\documentclass{aastex}
\makeatletter
\let\@dates\relax
\makeatother

\usepackage[squaren, Gray, cdot]{SIunits}
\usepackage{color}
\usepackage[hyphens]{url}
\usepackage{psfrag}
\usepackage{soul}

\newcommand{\alphamag}{\alpha_{\rm mag}}
\newcommand{\betaobs}{\beta_{\rm obs}}

\newbox\grsign \setbox\grsign=\hbox{$>$} \newdimen\grdimen \grdimen=\ht\grsign
\newbox\simlessbox \newbox\simgreatbox \newbox\simpropbox
\setbox\simgreatbox=\hbox{\raise.5ex\hbox{$>$}\llap
     {\lower.5ex\hbox{$\sim$}}}\ht1=\grdimen\dp1=0pt
\setbox\simlessbox=\hbox{\raise.5ex\hbox{$<$}\llap
     {\lower.5ex\hbox{$\sim$}}}\ht2=\grdimen\dp2=0pt
\setbox\simpropbox=\hbox{\raise.5ex\hbox{$\propto$}\llap
     {\lower.5ex\hbox{$\sim$}}}\ht2=\grdimen\dp2=0pt

\shorttitle{NuSTAR Observations of the Anomalous X-Ray Pulsar 1E 2259+586}
\shortauthors{Vogel et al.}

\begin{document}

\title{\emph{NuSTAR} Observations of the Magnetar 1E 2259+586}

\author{Julia K. Vogel\altaffilmark{1,2}, Romain Hasco\"{e}t\altaffilmark{3},Victoria M. Kaspi\altaffilmark{4}, Hongjun An\altaffilmark{4}, Robert Archibald\altaffilmark{4}, Andrei M. Beloborodov\altaffilmark{3}, Steven E. Boggs\altaffilmark{5}, Finn E. Christensen\altaffilmark{6}, William W. Craig\altaffilmark{2,5}, Eric V. Gotthelf\altaffilmark{3}, Brian W. Grefenstette\altaffilmark{7}, Charles J. Hailey\altaffilmark{3}, Fiona A. Harrison\altaffilmark{7}, Jamie A. Kennea\altaffilmark{8}, Kristin K. Madsen\altaffilmark{7}, Michael J. Pivovaroff\altaffilmark{2}, Daniel Stern\altaffilmark{9}, William W. Zhang\altaffilmark{10}\\}
\altaffiltext{1}{Corresponding author, vogel9@llnl.gov}
\altaffiltext{2}{Physics Division, Physical and Life Sciences Directorate, Lawrence Livermore National Laboratory, Livermore, CA 94550, USA}
\altaffiltext{3}{Physics Department and Columbia Astrophysics Laboratory, Columbia University, New York, NY 10027, USA}
\altaffiltext{4}{Department of Physics, McGill University, Montreal, Quebec H3A 2T8, Canada}
\altaffiltext{5}{Space Sciences Laboratory, University of California, Berkeley, CA 94720, USA}
\altaffiltext{6}{DTU Space, National Space Institute, Technical University of Denmark, Elektrovej 327, DK-2800 Lyngby, Denmark}
\altaffiltext{7}{Cahill Center for Astronomy and Astrophysics, California Institute of Technology, Pasadena, CA 91125, USA}
\altaffiltext{8}{Department of Astronomy and Astrophysics, Pennsylvania State University, University Park, PA 16802, USA}
\altaffiltext{9}{Jet Propulsion Laboratory, California Institute of Technology, Pasadena, CA 91109, USA}
\altaffiltext{10}{Goddard Space Flight Center, Greenbelt, MD 20771, USA}

\begin{abstract}
We report on new broad band spectral and temporal observations of the magnetar 1E 2259+586, which is located in the supernova remnant CTB 109. Our data were obtained simultaneously with the \emph{Nuclear Spectroscopic Telescope Array (NuSTAR)} and \emph{Swift}, and cover the energy range from $0.5-79$~keV. We present pulse profiles in various energy bands and compare them to previous \emph{RXTE} results. The \emph{NuSTAR} data show pulsations above $20$~keV for the first time and we report evidence that one of the pulses in the double-peaked pulse profile shifts position with energy. The pulsed fraction of the magnetar is shown to increase strongly with energy. Our spectral analysis reveals that the soft X-ray spectrum is well characterized by an absorbed double-blackbody or blackbody plus power-law model in agreement with previous reports. Our new hard X-ray data, however, suggests that an additional component, such as a power-law, is needed to describe the \emph{NuSTAR} and \emph{Swift} spectrum. We also fit the data with the recently developed coronal outflow model by Beloborodov (2013a) for hard X-ray emission from magnetars. The outflow from a ring on the magnetar surface is statistically preferred over outflow from a polar cap.
\end{abstract}


\section{Introduction}

Magnetars are a class of young, isolated neutron stars (NSs). Their defining features are their sporadic outbursts in X-rays and soft gamma-rays along with luminosities often much larger than their expected spin-down powered X-ray emission. To date, $21$ confirmed magnetars and $5$ candidates have been reported\footnote{\url{http://www.physics.mcgill.ca/~pulsar/magnetar/main.html}}~\cite{Olausen2014}. Spin periods for magnetars lie in the narrow range of $P=2.1-11.8$~s with period derivatives of the order of $\dot{P}=10^{-11}$~s~s$^{-1}$. The characteristic ages of these objects ($\tau= P/2\dot{P}$) imply that magnetars are rather young neutron stars with ages of the order of several thousand years. However, especially for young objects, the spindown-inferred age is only a crude estimate. The young ages are, however, generally supported for most magnetars by their specific locations as well as their association with supernova remnants (SNRs). Both magnetar outbursts and a substantial fraction of their persistent X-ray emission are believed to be powered by their intense magnetic fields~\cite{Thompson1993,Thompson1996}, with strengths of $10^{14}-10^{15}$~G inferred assuming magnetic dipole braking in vacuum ($B\propto \sqrt{P\dot{P}}$).

The magnetar model was originally formulated to explain the behavior of a sub-class of magnetars, the so-called Soft Gamma Repeaters (SGRs), the most active of these objects. However, observations of similar behavior such as bursting activity~\cite{Gavriil2002,Kaspi2003} for Anomalous X-ray Pulsars (AXPs), another previously identified magnetar sub-class, indicate that SGRs and AXPs are likely to be at different ends of a continuous magnetar activity spectrum.

\begin{figure}[t!]
\begin{centering}
\includegraphics[width=0.5\textwidth]{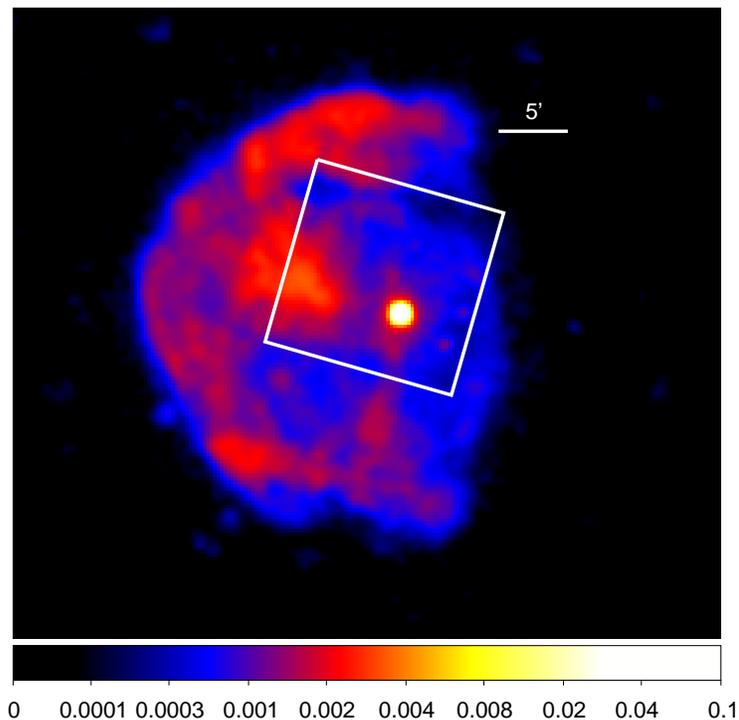}
\caption{\emph{ROSAT} PSPC image of SNR CTB 109 ($0.1-2.4$~keV). The white frame indicates the \emph{NuSTAR} field of view (FoV), but no corresponding data have been included here. The bright point source is the magnetar 1E 2259+586. }
\label{fig:CTB109}
\end{centering}
\end{figure}

One of the most studied magnetars to date is 1E 2259+586. It was the first AXP to be discovered~\cite{Fahlman1981,Fahlman1983} and also played a key role in the unification of these neutron stars with very high magnetic fields, when it exhibited a series of over $80$ X-ray bursts in 2002~\cite{Kaspi2003}. This AXP has shown a very stable spin-down rate~\cite{Kaspi1999} as well as pulsed flux emission apart from glitches\footnote{A glitch is a sudden spin-up of a neutron star. In one case a sudden spin-down has been observed~\cite{Archibald2013}, which is referred to as an anti-glitch.} in 2002 and 2007~\cite{Icdem2012,Dib2014} and a rare anti-glitch in 2012~\cite{Archibald2013}. 1E 2259+586 is located near the center of SNR CTB 109 (G$109.1-1.0$), which is known for its half-shell morphology in both radio and X-rays (see Fig.~\ref{fig:CTB109}). It is at a distance of $4.0\pm0.8$~kpc~\cite{Tian2010}.
%
%
The spin period of 1E 2259+586 is $P=6.98$~s with a spin-down rate of $\dot{P}=0.05\times 10^{-11}$~s s$^{-1}$. This implies a surface dipole magnetic field of $0.59\times 10^{14}$~G, which is towards the lower end of the typical range of magnetar magnetic fields, and the characteristic age is estimated to be $230$~kyr. While no radio counterpart at the position of the X-ray pulsar has been identified down to a level of $7~\mu$Jy~\cite{Archibald2013}, persistent emission has been detected from the mid- to far-infrared, all the way up to X-rays (see, e.g., Kuiper et al. 2006 for a brief review of previous observations). The soft X-ray spectrum is typical of AXPs and, in the $0.5-7.0$~keV band, was described by Patel et al. (2001) as a blackbody spectrum with temperature $kT=0.412(6)$~keV and a soft power-law with $\Gamma=3.6(1)$ for $N_{\rm H}=0.93(3)\times 10^{22}$~cm$^{-2}$. Zhu et al. (2008) updated these values in the $0.6-12.0$~keV band with \emph{XMM-Newton} data taken post-outburst in 2002, yielding $kT=0.400(7)$~keV and $\Gamma=3.75(3)$ for $N_{\rm H}=1.012(7)\times 10^{22}$~cm$^{-2}$. It is interesting that significant spectral changes were observed by Woods et al. (2004) at energies $<10$~keV when comparing pre- and post-burst data for the 2002 outburst. Timing and spectral properties for hard X-ray emission ($8-24$~keV) from 1E 2259+586 were first derived by Kuiper et al. (2006) using \emph{RXTE} data. They also compared their results for the pulsed spectrum to upper limits from \emph{IBIS} ISGRI and \emph{COMPTEL}~\cite{denHartog2006}. For the \emph{RXTE} data, an absorbed double power-law model yields a good fit to the pulsed spectrum with soft and hard power-law indices of $\Gamma_{1}=4.26(1)$ and $\Gamma_{2}=-1.02^{+0.24}_{-0.13} $ for $N_{\rm H}=0.93(3)\times 10^{22}$~cm$^{-2}$. Therefore, this AXP follows others in exhibiting an onset of dramatic hardening above $10$~keV. For 1E 2259+586, Kuiper et al. (2006) found that the power-law components become equally strong at $E=15.8\pm 2.3$~keV. For energies beyond 20 keV, conclusive confirmation of the dramatic hardening for 1E 2259+586 has not been available until now. 

In this paper, we report on the spectral and temporal properties
of the magnetar 1E 2259+586 in the $0.5-79$ keV band. Our analysis is based on data acquired with the \emph{Nuclear Spectroscopic Telescope Array }(\emph{NuSTAR}) and the \emph{Swift} X-Ray Telescope (XRT).
In Section 2, we describe the observations, and following this we present the results of our data analysis in Section 3. We apply the coronal outflow model of Beloborodov (2013a) to the hard X-ray emission data, and show that the model is consistent with the phase-resolved spectra. A discussion of our results can be found in Section 4, and in Section 5 we conclude and summarize our findings.

\section{Instruments and Observations}
\emph{NuSTAR} is the first hard X-ray focusing telescope in space with sensitivity in the energy range of $3-79$~keV~\cite{Harrison2013}. The instrument consists of two co-aligned focusing optics~\cite{Hailey2010} with CdZnTe detectors in the focal plane~\cite{Harrison2010}. Each focal-plane module consists of four detector chips and the two telescope modules are dubbed FPMA and FPMB. \emph{NuSTAR}'s energy resolution is $400$~eV at $10$~keV (Full Width at Half Maximum, FWHM) and the observatory provides an angular resolution of $58\arcsec$ HPD\footnote{Half-Power Diameter} ($18\arcsec$ FWHM). Its temporal resolution of $2~\mu$s  is more than adequate for studying the $6.98$~s AXP 1E 2259+586. For more information on \emph{NuSTAR}, see Harrison et al. (2013).

\emph{Swift}~\cite{Gehrels2004} is a multi-wavelength mission dedicated to the study of gamma-ray bursts. The \emph{Swift} XRT~\cite{Burrows2005} is one of three instruments flown on the satellite. It is a Wolter-I type optic with a Charge-Coupled Device (CCD) focal plane detector. The instrument is sensitive in the energy range from $0.2$ to $10$~keV with an energy resolution of $140$~eV at $5.9$~keV and a point spread function of $22\arcsec$ (HPD) at $8.1$ keV. 

The \emph{NuSTAR} observations of 1E 2259+586 began on $24$ April $2013$ at UT 21:51:07 with the last of four observations concluding on 18 May $2013$ at UT 06:01:07. The total net exposure for the combination of all four observations is $157.4$~ks. A simultaneous \emph{Swift} observation (XRT, photon-counting mode) was carried out to extend the spectral coverage down to $\sim 0.5$~keV. While \emph{NuSTAR} is very sensitive in the hard X-ray band, the low energy data from the \emph{Swift} XRT ($0.2-10$~keV) improve our ability to constrain the thermal components which are softer and emit mostly below the \emph{NuSTAR} pass-band. The \emph{Swift} observation started at UT 00:23:55 on $25$ April $2013$ and was split into three parts with a total exposure time of $29.9$~ks.

\begin{figure}[t!]
\begin{centering}
\includegraphics[width=0.95\textwidth]{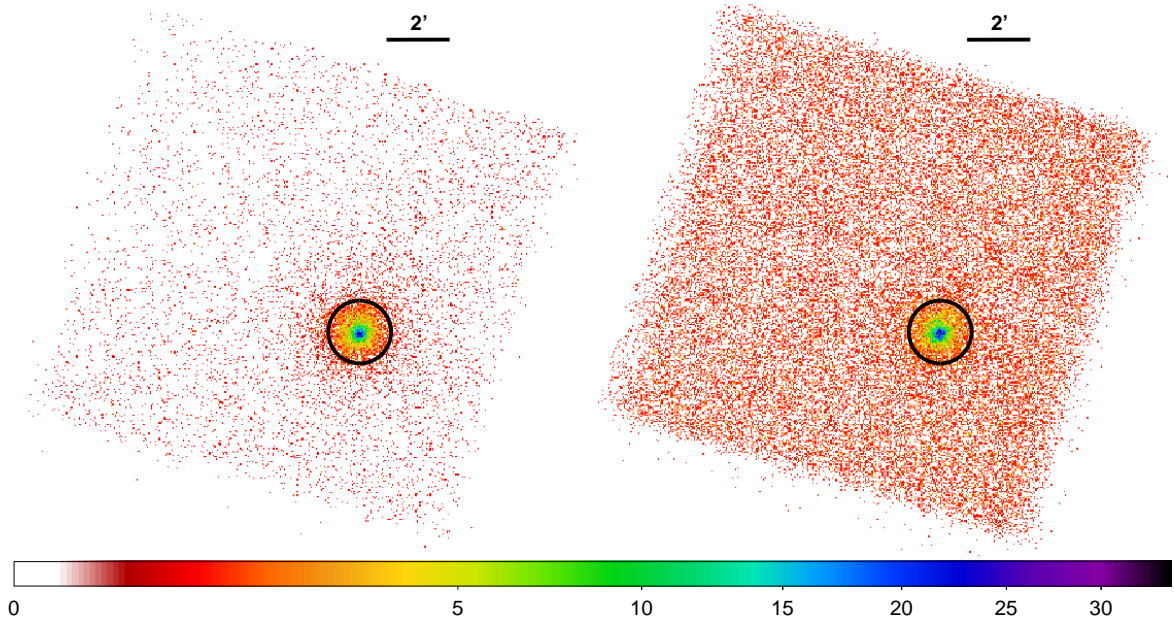}
\caption{Images of unsmoothed, exposure-corrected \emph{NuSTAR} data in two energy ranges (Left: $3-4$~keV, right: $4-79$~keV) with comparable numbers of counts in the signal extraction region of $60\arcsec$ radius.}
\label{fig:Nustardata}
\end{centering}
\end{figure}

The \emph{NuSTAR} data were processed using the standard  \emph{NuSTAR} Data Analysis Software {\tt NuSTAR\-DAS} version 1.2.0 along with CALDB version 20131007 and HEASOFT version 6.13 as available on HEASARC\footnote{See \url{http://heasarc.gsfc.nasa.gov}}. For the \emph{Swift} data, cleaned event files were produced using {\tt xrtpipeline} along with the HEASARC remote CALDB\footnote{See \url{http://heasarc.nasa.gov/docs/heasarc/caldb/caldb_remote_access.html}} employing the standard filtering procedure of Capalbi et al. (2005). Further processing steps applied to these files are outlined below. Details about the different observations are shown in Table~\ref{tab:observations}, which summarizes all data sets used in this analysis. Figure~\ref{fig:Nustardata} shows the unsmoothed, exposure-corrected \emph{NuSTAR} image of 1E 2259+586 in two energy bands, chosen to yield a comparable number of counts in the signal extraction region of radius $60\arcsec$.
%
%
%
%
%
\begin{table}
\begin{center}
\caption{\emph{NuSTAR} and \emph{Swift} observations of 1E 2259+586 used in this study.\label{tab:observations}}
\vspace{4mm}
\begin{tabular}{cccccc}
\tableline\tableline
Observatory & Mode& ObsID & Date [MJD]\tablenotemark{a} & Date & Exp [ks]  \\
\tableline
\emph{NuSTAR} & $\cdots$ & 30001026002 & 56406 & April 24, 2013 & 37.3 \\
\emph{NuSTAR} & $\cdots$ & 30001026003 & 56407 & April 25, 2013 & 15.4 \\
\emph{NuSTAR} & $\cdots$ & 30001026005 & 56408 & April 26, 2013 & 16.3 \\
\emph{NuSTAR} & $\cdots$ & 30001026007 & 56428 & May 16, 2013 & 88.4 \\
\emph{Swift} & PC & 00080292002 & 56407 & April 25, 2013 &  13.1\\
\emph{Swift} & PC & 00080292003 & 56408 & April 26, 2013 & 14.1 \\
\emph{Swift} & PC & 00080292004 & 56410 & April 28, 2013 & 2.7 \\
\tableline
\end{tabular}
\tablenotetext{a}{At the start of data acquisition.}
\end{center}
\end{table}
%
\section{Analysis and Results}
%
%
\subsection{Timing Analysis}
\label{sec:Timing}
For the timing analysis, we extracted photons in a circular region of $60\arcsec$ radius around the nominal source position for the \emph{NuSTAR} data, while for \emph{Swift} we used a $22\arcsec$ radius. The energy bands used for spectral analysis were $5-79$~keV and $0.5-10$~keV for \emph{NuSTAR} and \emph{Swift}, respectively. Since 1E 2259+586 is located in CTB 109, the SNR background must be considered. Sasaki et al. (2004) pointed out that no emission above $4$~keV was observed in the \emph{Chandra} data apart from the pulsar and the \emph{NuSTAR} data agrees with this. Since in one of the two \emph{NuSTAR} modules (FPMA) stray light is observable at the edge of the field of view, we excluded this region when defining the background extraction area. For \emph{NuSTAR}, background was extracted from a circular region of radius $100\arcsec$ nearby the source region. We also compared our results using these background regions to those obtained using the mission's background model {\tt nuskybgd}~\cite{Wik2014}, and confirmed that all results were compatible within the corresponding uncertainties.

Due to the good spatial resolution of the \emph{Swift} XRT, background could be extracted from an annular region around the neutron star with an inner radius of $80\arcsec$ and an outer radius of $200\arcsec$. Before continuing we checked if any pile-up events were present in the data. We followed the standard \emph{Swift} procedure\footnote{See \url{http://www.swift.ac.uk/analysis/xrt/pileup.php}} and determined the count rate in a circular region of $47.2\arcsec$ radius ($20$ pixels) centered on the source to be above $0.5$~ct/s, indicating that the central 1-2 pixel radius region of the bright core might suffer from pile-up. We removed this area for the subsequent analysis by substituting the circular source extraction region by an annulus with inner and outer radii of $5\arcsec$ and $22\arcsec$, respectively. We also re-generated the ARFs to correct the flux of the spectrum for the loss of counts due to the use of an annular region as well as the correction factor for the light curves\footnote{See \url{http://www.swift.ac.uk/analysis/xrt/arfs.php}}.

In the next step, we applied a barycentric correction to the selected source events using the multi-mission tool {\tt barycorr} with the corresponding orbital and clock correction files at the \emph{Chandra} position of $\alpha=23^{h}01^{m}08^{s}.295$ and $\delta=+58\arcdeg52\arcmin44\arcsec.45$ (J2000.0) reported by Patel et al. (2001). Following the H-test method~\cite{deJager1989}, we searched for pulsations and determined the best period for the significant pulsations to be $P($\emph{NuSTAR}$)=6.97914(2)$~s and $P($\emph{Swift}$)=6.97915(2)$~s. The measured periods were in agreement with those obtained from the ephemeris of the \emph{Swift} monitoring program of 1E 2259+586 (see Table~\ref{tab:ephemeris}).
%
\begin{table}
\begin{center}
\caption{\emph{Swift} Ephemeris for 1E 2259+586 ~\cite{Archibald2013}.\label{tab:ephemeris}}
\vspace{4mm}
\begin{tabular}{cc}
\tableline\tableline
Parameter& Value  \\
\tableline
Spin frequency $f_{0}$ [s$^{-1}$] & $0.14328414235(81)$\\
Spin frequency derivative $\dot{\nu}_{0}$ [s$^{-2}$] & $-10.14(12)\times 10^{-15}$\\
Epoch & MJD $56250.000$\\
Start observing epoch & MJD $56201.284$\\
Stop observing epoch & MJD $56439.246$\\
\tableline
\end{tabular}
\end{center}
\end{table}
%

\begin{figure}[t!]
\centering
\includegraphics[width=0.95\textwidth]{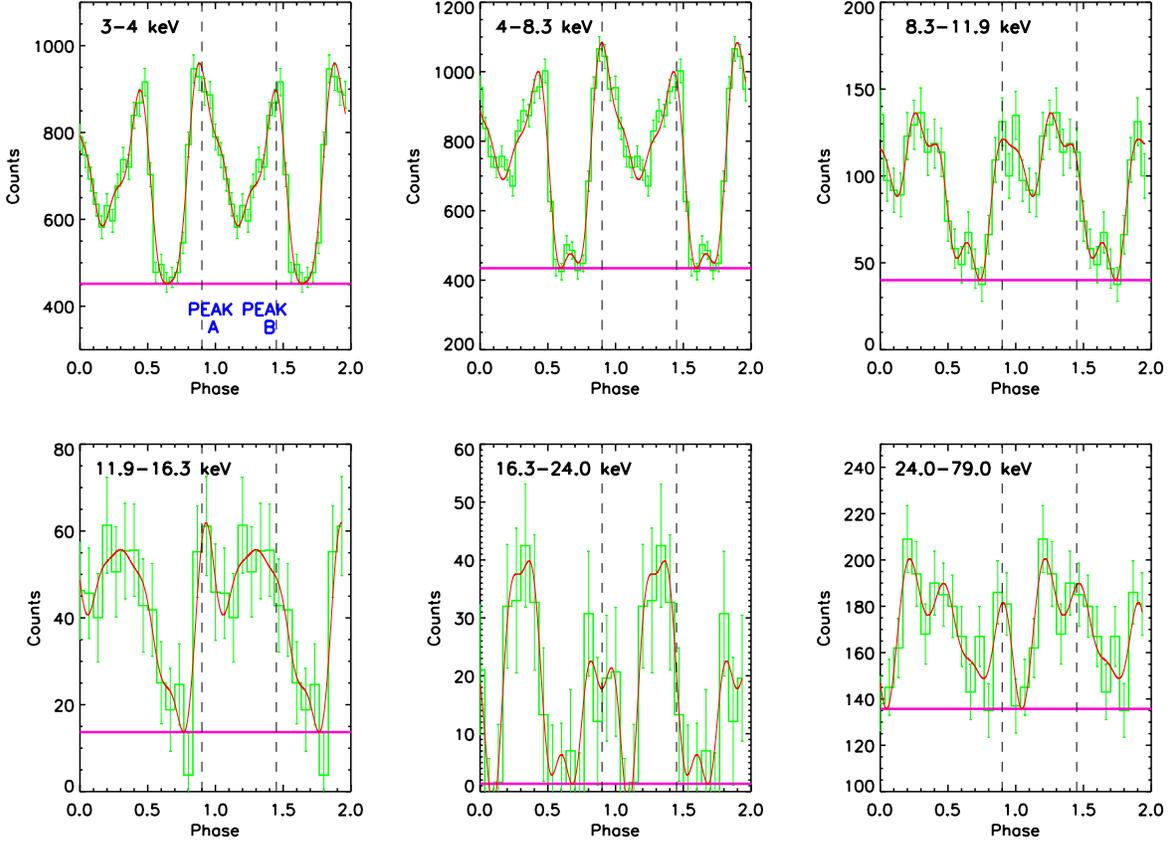}
\caption{Pulse profiles for 1E 2259+586 from \emph{NuSTAR} data in various energy bands. All but the pulse profile in the band from $24-79$~keV, which has limited statistics, are background-subtracted. Energy bands up to $24$~keV were chosen to enable direct comparison with Kuiper et al. (2006). The dashed lines near phase $0.9$ and $1.45$ are meant to serve as a guide to the eye to compare alignment of peak A and peak B. Note that the y-axis labels differ and the zero is suppressed in some of the plots. Also shown is the Fourier representation of the pulse profiles (red) used to estimate the pulsed emission level, denoted by the horizontal line. }
\label{fig:pulseprofile}
\end{figure}

The resulting pulse profiles are shown in Figure~\ref{fig:pulseprofile} for six energy ranges: $3$-$4$~keV, $4$-$8.3$~keV, $8.3$-$11.9$~keV, $11.9$-$16.3$~keV, $16.3$-$24.0$~keV, $24.0$-$79.0$~keV. The bands up to $24.0$~keV were chosen to coincide with those reported by Kuiper et al. (2006) for direct comparison. All pulse profiles, with the exception of the highest energy band ($24-79$~keV), are background-subtracted. The statistical significance of the observed pulsations in all energy bands is larger than $99\%$ (p-value\footnote{The p-value is the probability of obtaining a test statistic which is at least as extreme as the one that was actually observed under the assumption that the null hypothesis (in our case, no pulsation) is true.} smaller than $0.01$). For the lower five energy bands, the pulse profiles agree generally with those reported by Kuiper et al. (2006); however, in the energy ranges between $11.9$ and $24.0$~keV slight differences are apparent. In the highest energy band ($24-79$~keV), significant pulsations are observable for the first time. We note, however, that due to limited statistics we did not subtract the background in order to avoid negative bin counts. The nominal background count rate in this case is $6.56\times10^{-4}$ cts$/$s.
%
%

As previously pointed out by Kuiper et al. (2006), there is a gradual change in pulse morphology observable with increasing energy. The double-peak profile at low energies (below $\sim 8$~keV) evolves into a less pronounced double-peak or possibly a single peak profile for intermediate energies (about $8-16$~keV) and back to a double-peak structure at higher energies. A possible way to describe this development is that while for energies below $\sim 8$~keV, the pulse near phase $0.9$ (`Peak A') tends to dominate, it becomes less significant for increasing energy with respect to the second peak (`Peak B') near phase $1.45$. In the intermediate energy range  ($8-16$~keV), both peaks are approximately equally pronounced, while peak B starts to slightly dominate at energies above $\sim 16$~keV. A difference between \emph{NuSTAR} and \emph{RXTE} profiles is apparent in the energy bands of $11.9-16.3$~keV and $16.3-24.0$~keV. While Kuiper et al. (2006) observe a clearly dominating peak B in this energy range, our analysis shows only a small lead for this pulse over peak A.

\begin{figure}[ht!]
\centering
\includegraphics[width=0.95\textwidth]{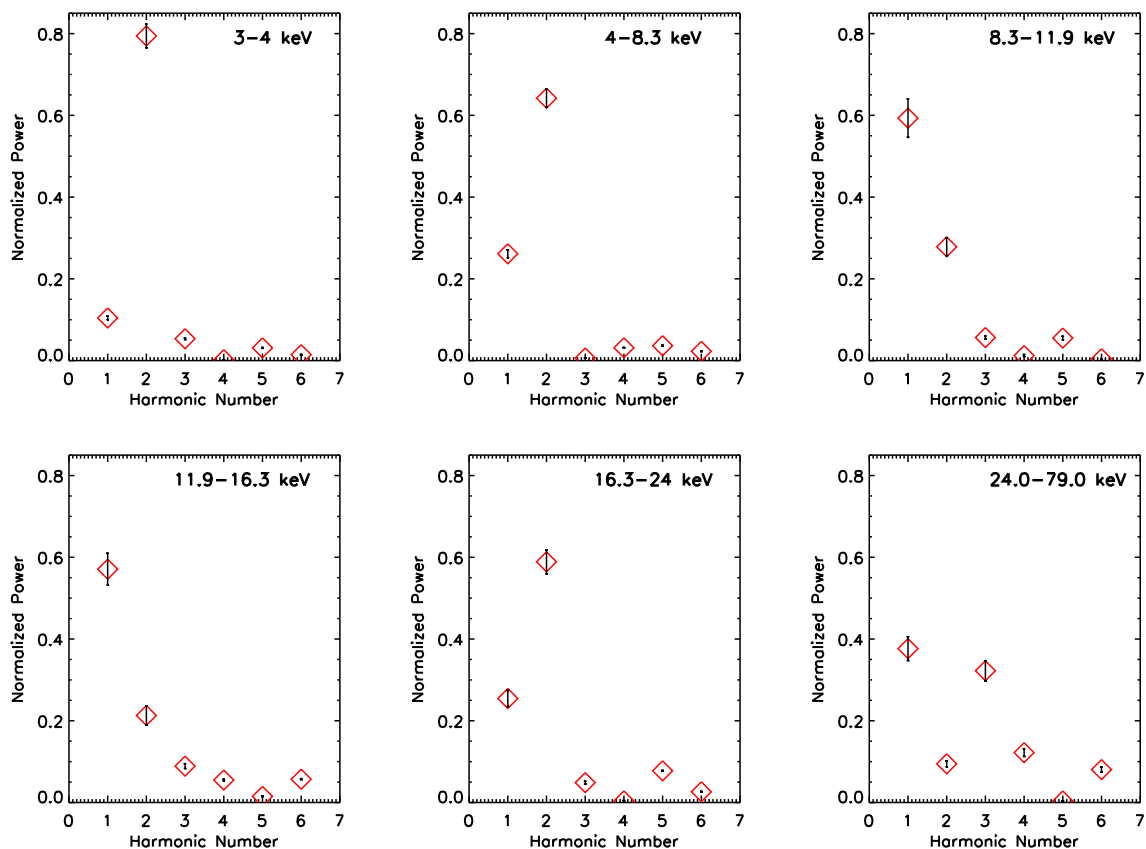}
\caption{Evolution of the Fourier power harmonic distribution of the pulse profile for 1E 2259+586 as a function of energy. The energy bands shown correspond to those displayed in Figure~\ref{fig:pulseprofile}. Power levels have been normalized to the total power of the first six harmonics. Note that the error bars for the $24-79$~keV band are artificially small, since the corresponding pulse profile has not been background-subtracted. Errors will significantly increase in this energy band (due to low statistics) if background is taken into consideration.}
\label{fig:fourier}
\end{figure}

In order to quantify the energy dependence of the pulse morphology, we decompose the pulse profiles in terms of their Fourier components. This decomposition is shown in Figure~\ref{fig:fourier}. 
%
%
We find six harmonics to be sufficient, and we included these in the plots. We observe that the ratio of the second harmonic to the first harmonic changes with energy. For energies below about $8$~keV, the second harmonic dominates over the first, while in the intermediate range of $8-16$~keV the first harmonic takes the lead, before the second harmonic dominates again for $16-24$~keV. If the main contribution originates from the first harmonic number, this indicates that there is either one single peak in the pulse profile or a double peak with maxima of similar magnitude, while a leading second harmonic number points towards a double peak with pulses of different height and/or shape. This Fourier decomposition therefore reflects the change in dominance of one peak over the other in the double-peak morphology or, likewise, a possible change from a 2-peak structure to one peak and back. For the \emph{NuSTAR} band ($E>24$~keV), the first and third harmonics dominate ($2$, $4$, and $6$ contribute at the $10\%$ level). When interpreting this power spectrum (lower right plot of Figure~\ref{fig:fourier}) one should keep in mind that background was not subtracted and the peak-to-valley ratio is not very large. Here the first and (small) second harmonic take care of the wide peak B (from phase $1-1.8$, including the small feature in the main minumum), while the third harmonic describes the narrower, subdominant peak A.

As previously reported by Kuiper et al. (2006) as well as Gavriil \& Kaspi (2002), a small pulse-like feature can be observed in the main minimum around phase $0.60-0.65$. This feature is the main reason for the presence of higher orders of harmonics ($n\geq3$) in the Fourier decomposition (see Fig.~\ref{fig:fourier}) of the pulse profile.

We also note that there is a hint of a small feature in the intermediate minimum between peak A and B around phase $1.2$, best seen in the lowest two energy bands. With the current statistics, however, we do not consider this feature to be significant.

Another observation is a possible shift of the peak B position in phase towards lower values with increasing energy. Dashed lines near phase $0.9$ and $1.45$ have been added to Figure~\ref{fig:pulseprofile} to guide the eye for alignment comparison of peak A and peak B. In order to test this pulse-shift hypothesis we determined the mean value of the peak position for each peak in various energy bands by considering the unbinned group phase between phase $0.75-1.1$ for peak A and $1.2-1.55$ for peak B. In addition, we crosschecked our results by fitting a Gaussian to the peaks. The results for the mean and its error are listed in Table~\ref{tab:pulseshift} and indicate that peak A remains at the same phase position of $0.937(17)$ throughout all energy ranges ($0.933(3)$ for $E\lesssim 8$~keV, $0.940(16)$ for $E\gtrsim 8$~keV). 
%
\begin{table}[b!]
\begin{center}
\caption{Mean peak position for peak A and B in various energy bands.\label{tab:pulseshift}}
\vspace{4mm}
\begin{tabular}{ccccc}
\tableline\tableline
Energy band                  & Position & Error    & Position & Error    \\
$\lbrack$keV$\rbrack$ & peak A   & peak A & peak B   & peak B \\
\tableline
 $  3.0-  4.0$ &     $0.93$ &   $<0.01$  &   $1.39$ &   $<0.01$  \\
 $  4.0-  8.3$ &     $0.94$ &   $<0.01$  &   $1.38$ &   $<0.01$  \\
 $  8.3-11.9$ &     $0.94$ &   $0.01$    &   $1.35$ &   $0.01$     \\
 $11.9-16.3$ &     $0.94$ &   $0.01$    &   $1.35$ &   $0.01$     \\
 $16.3-24.0$ &     $0.93$ &   $0.01$    &   $1.36$ &   $0.01$     \\
\tableline
\end{tabular}
\end{center}
\end{table}
%
For peak B, there is a slight tendency to change position towards lower phase values for higher energies. Below $\sim 8$~keV, peak B can be found at $1.384(3)$, above these energies it is located at $1.349(17)$. We note that the shift is not very significant at the given statistics, and more data are needed to confirm or rule out the observed tendency.

\begin{figure}[ht!]
\begin{centering}
\includegraphics[width=0.65\textwidth]{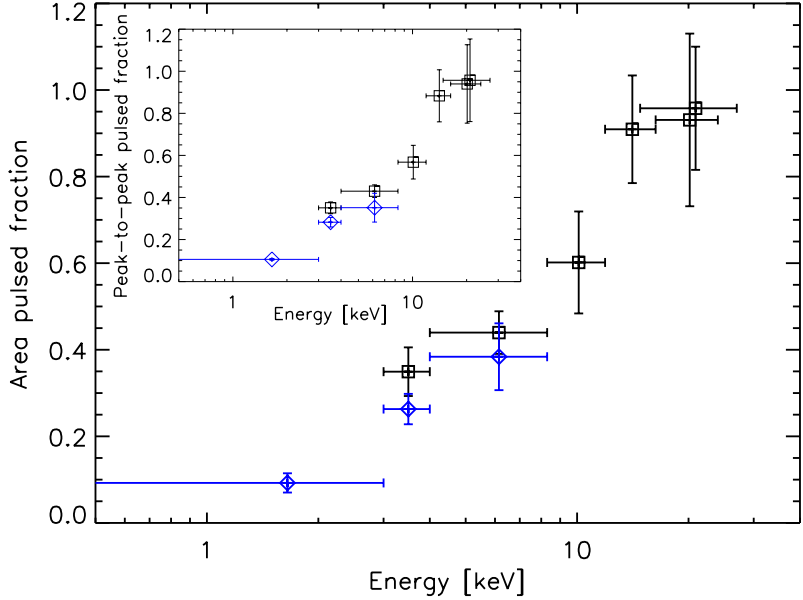}
\caption{Area pulsed fraction as function of energy for 1E 2259+586 from \emph{NuSTAR} (black squares) and \emph{Swift} data (blue diamonds). The energy bands correspond to those presented in Kuiper et al. (2006) for the \emph{RXTE} PCA and HEXTE data. The small inset plot shows the conventionally defined $PF_{\rm P2P}$. In both cases the pulsed fraction strongly increases with energy.}
\label{fig:pulsedfraction}
\end{centering}
\end{figure}

In addition, we used the ephemeris-folded profiles to determine the pulsed fraction (PF) as a function of energy for our observations. A pulse profile that changes with energy or time makes it challenging to determine the pulsed fraction of the source accurately and different methods are commonly used in the literature (see e.g. Archibald et al. 2007, Archibald et al. 2014 for a discussion). All of them have their advantages and disadvantages and therefore there are some caveats to keep in mind. The area pulsed fraction is probably the most natural and physically meaningful definition. It is defined as the difference between the pulsed flux and the constant flux integrated over a full phase cycle and can be calculated according to 
 %
\begin{equation}
PF_{\rm area}=\frac{\frac{1}{N}\displaystyle\sum\limits_{j=1}^{N} p_{j}-p_{\rm min}}{\frac{1}{N}\displaystyle\sum\limits_{j=1}^{N} p_{j}},
\end{equation}
where $p_{j}$ is the number of events in the $j$-th phase bin, $N$ is the total number of bins, and $p_{\rm min}$ is the minimum flux. Note that the determination of the true minimum and its error is the main challenge and the biggest caveat when using $PF_{\rm area}$, because both noise and binning show a tendency to bias the values of the area pulsed fraction upwards. In  Figure~\ref{fig:pulseprofile} we included the results for $p_{\rm min}$ as the minimum off-pulse value determined from the Fourier representation (horizontal magenta line). The results for the area pulsed fraction using the $p_{\rm min}$ determined in this way are shown in Figure~\ref{fig:pulsedfraction}. Note that the \emph{NuSTAR} energy bands here (black squares) correspond to those presented in Kuiper et al. (2006) for the \emph{RXTE} PCA (five bands in the range from $3.0-24.0$~keV) and HEXTE data ($14.8-27.0$~keV). The \emph{Swift} data (blue diamonds) covers the two lowest \emph{RXTE} PCA bands in Kuiper et al. (2006) plus an additional band from $0.3-3.0$~keV. Background was subtracted in all energy bands. 

The small inset plot in the same Figure displays the standard pulsed fraction (peak-to-peak pulsed fraction), conventionally defined as ratio of peak flux minus minimum flux to the total flux
%
\begin{equation}
PF_{\rm P2P}=\frac{F_{\rm max}-F_{\rm min}}{F_{\rm max}+F_{\rm min}}.
\end{equation}
This quantity is extensively used in the literature since it is straightforward to calculate, although the difficulty in determining the maximum and minimum flux values accurately is biased in a way similar to the area pulsed fraction. Furthermore, the $PF_{\rm P2P}$ cannot provide information on the total energy since the peak width does not feed into the flux variation. The $PF_{\rm area}$ and $PF_{\rm P2P}$  derived from our data agree with each other within uncertainties and show the same trend: the pulsed fraction increases dramatically with increasing energy. For energies around $10$~keV, $PF_{\rm area}$ is $60\%\pm12\%$, while for energies around $20$~keV the fraction reaches $96\%\pm14\%$. 

%
\begin{figure}[ht!]
\begin{centering}
\includegraphics[width=0.65\textwidth]{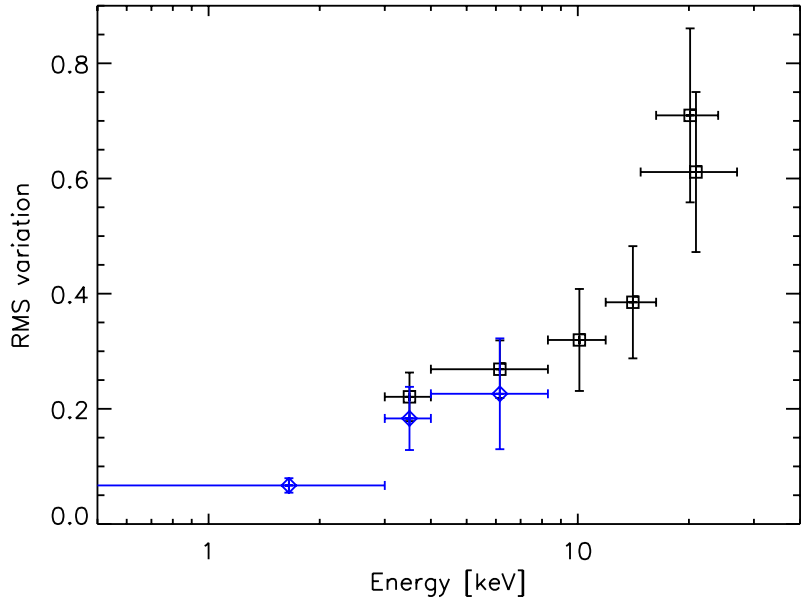}
\caption{RMS deviation of the flux from the mean as function of energy for 1E 2259+586 from \emph{NuSTAR} (black squares) and \emph{Swift} data (blue diamonds). The energy bands correspond to those presented in Kuiper et al. (2006) for the \emph{RXTE} PCA and HEXTE data. Similarly to the area and peak-to-peak pulsed fraction also the $PF_{\rm RMS}$ strongly increases with energy, even though the absolute values differ due to the definition of the RMS estimator.}
\label{fig:rmspulsedfraction}
\end{centering}
\end{figure}

%
Following the method presented in Gonzalez et al. (2010) (see also the forthcoming publication by Archibald et al. 2014 for details), we additionally used a root-mean-square (RMS) estimator as a measure for the deviation of the pulsed flux from its mean defined as
%
\begin{equation}
PF_{\rm RMS}=\frac{ \sqrt{2\displaystyle\sum\limits_{k=1}^{6}\left(\left(a_{k}^{2}+b_{k}^{2}\right)-\left(\sigma_{a_{k}}^{2}+\sigma_{b_{k}}^{2}\right)\right)}}
{a_{0}}
\label{eq:rmspf}
\end{equation}
%
where the Fourier coefficients $a_{k}$ and $b_{k}$ are
%
\begin{equation}
a_{k}=\frac{1}{N}\displaystyle\sum\limits_{j=1}^{N}p_{j}\cos\left(\frac{2\pi kj}{N}\right),
\end{equation}
%
%
\begin{equation}
b_{k}=\frac{1}{N}\displaystyle\sum\limits_{j=1}^{N}p_{j}\sin\left(\frac{2\pi kj}{N}\right)
\end{equation}
%
and $\sigma_{a_{k}}$, $\sigma_{b_{k}}$ are the uncertainties of $a_{k}$ and $b_{k}$, respectively,
\begin{equation}
\sigma_{a_{k}}^{2}=\frac{1}{N^{2}}\displaystyle\sum\limits_{j=1}^{N}\sigma_{p_{j}}^{2}\cos^{2}\left(\frac{2\pi kj}{N}\right),
\end{equation}
%
%
\begin{equation}
\sigma_{b_{k}}^{2}=\frac{1}{N^{2}}\displaystyle\sum\limits_{j=1}^{N}\sigma_{p_{j}}^{2}\sin^{2}\left(\frac{2\pi kj}{N}\right).
\end{equation}
$p_{j}$ is again the number of events in the $j$-th phase bin, $\sigma_{p_{j}}$ is the uncertainty of $p_{j}$, and $N$ is the total number of bins. The number of Fourier harmonics included is $n=6$. Note that $PF_{\rm RMS}$ is the RMS deviation of the flux from its average and can be defined in the Fourier domain. It does not cover the full range of values from $0-1$ without additional (pulse shape dependent) renormalization, i.e. it is not a pulsed fraction in the conventional sense, even though it is often referred to as the RMS pulsed fraction in the literature. This definition yields a very robust measure for the pulsed power that is less sensitive to noise and therefore signficantly less biased than the area pulsed fraction and more meaningful than the peak-to-peak pulsed fraction. 
Figure~\ref{fig:rmspulsedfraction} shows the RMS variation for \emph{NuSTAR} (black squares) and \emph{Swift} data (blue diamonds). Note that the energy bands again correspond to those presented in Kuiper et al. (2006) for the \emph{RXTE} PCA (five bands in the range from $3.0-24.0$~keV) and HEXTE data ($14.8-27.0$~keV) and represent background-subtracted data in all energy bands. Since the definition of $PF_{\rm RMS}$ differs from those of $PF_{\rm area}$ and $PF_{\rm P2P}$ as mentioned above, $PF_{\rm RMS}$ does not have the same values as the other definitions in each energy band, but the same trend is observable in all cases: the pulsed fraction increases dramatically with increasing energy. 
Without renormalization we determined that the $PF_{\rm RMS}$ is $32\%\pm9\%$ for energies around $10$~keV, while for energies around $20$~keV it reaches $71\%\pm15\%$. 
%

%
%
%
\subsection{Spectral Analysis}
%
%
%
\subsubsection{Phase-Averaged Spectral Analysis}
\label{sec:phaseaveragedspectralana}
In this Section, we use the \emph{Swift} and \emph{NuSTAR} observations to derive a phase-averaged spectrum. The same extraction regions as defined in Section~\ref{sec:Timing} were used and pile-up for the \emph{Swift} data was corrected for as described above.

We started with fitting the \emph{Swift} data only in order to compare the results to previous soft X-ray measurements~\cite{Zhu08} as well as to obtain a best-fit $N_{\rm H}$ to be used for the combined of \emph{Swift} and \emph{NuSTAR} data sets. All extracted spectra were grouped to have at least $50$ counts per bin using the {\tt grppha} tool. We fit the resulting spectra with XSPEC v12.8.1 using an absorbed blackbody (BB) plus power-law (PL) model  ({\tt tbabs*(bbody+powerlaw)}) as well as an absorbed double blackbody ({\tt tbabs*(bbody+bbody)}). The \emph{Swift} spectrum is a little softer ($\Gamma=4.1(1)$) than previously reported ($\Gamma=3.75(3)$, Zhu et al. 2008), but consistent within uncertainties with these earlier results. Our best-fit result yields $N_{\rm H}=1.10(6)\times 10^{22}$~cm$^{-2}$ in the case of the blackbody plus power-law and  $N_{\rm H}=0.60(4)\times 10^{22}$~cm$^{-2}$ for the double-blackbody model. The first value agrees well with estimates of Durant \& van Kekwijk (2006) obtained from fitting individual absorption edges of O, Fe, Ne, Mg and Si in the \emph{XMM-Newton} RGS spectra. The latter value is in agreement with Zhu et al. (2008) and consistent with the best-fit $N_{\rm H}$-value for CTB 109, $N_{\rm H}=(0.5-0.7)\times 10^{22}$~cm$^{-2}$, as measured by Sasaki et al. (2004).

\begin{figure}[ht!]
\begin{centering}
\includegraphics[trim=0mm 1mm 0mm 0mm, clip=true,width=0.75\textwidth]{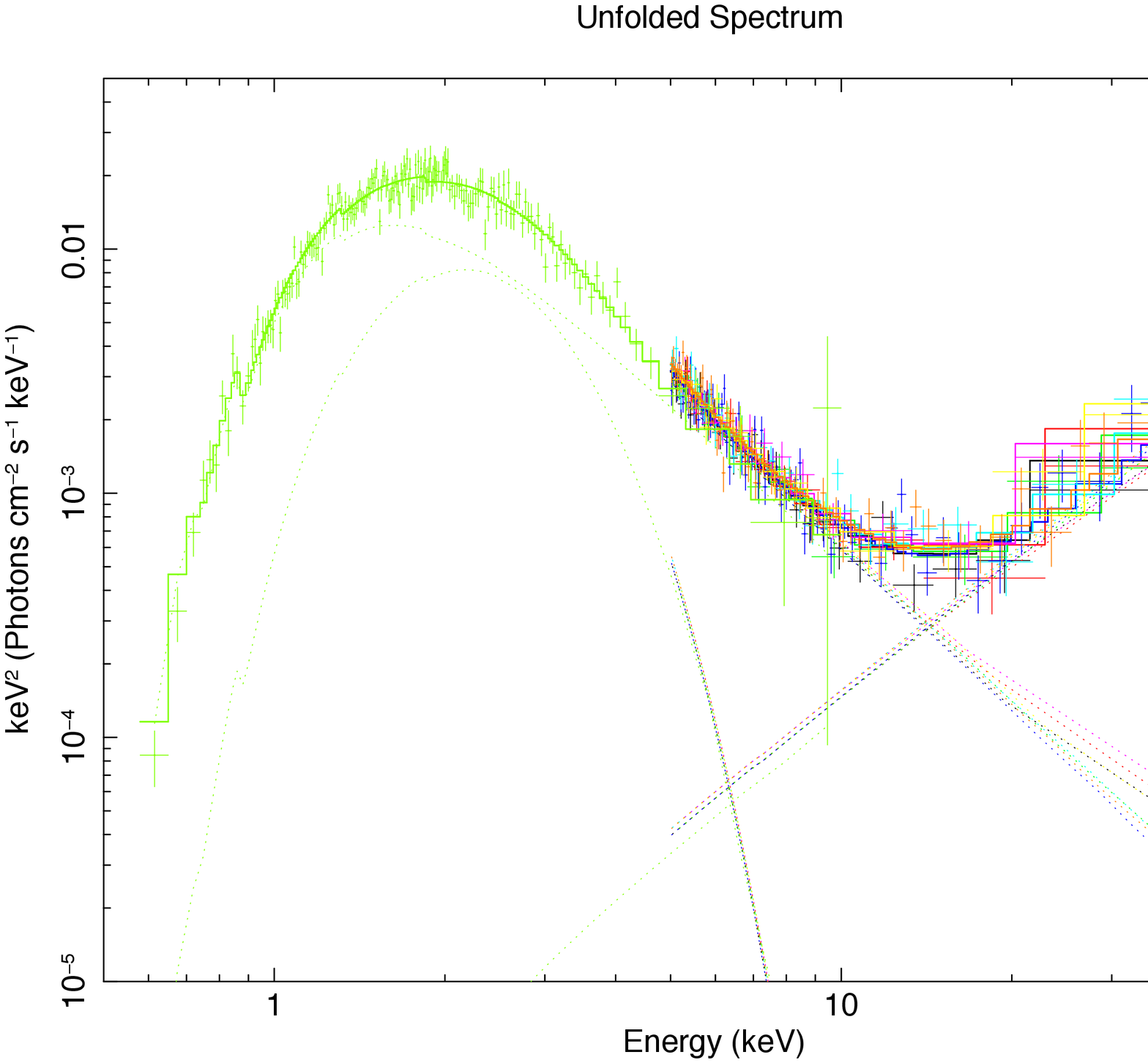}
\caption{Phase-averaged spectra of \emph{Swift} data (green) and all four \emph{NuSTAR} observations with the best-fit model (see Table~\ref{tab:spectralresults}). The additive components are included in the plot. The model shown is an absorbed blackbody plus a double power-law.}
\label{fig:bestfitspectra}
\end{centering}
\end{figure}

In a next step we fit the spectra from $0.5-79$~keV by using the \emph{NuSTAR} and \emph{Swift} data sets together. All model parameters were tied except for the cross-normalization factors, which were set to $1.0$ for \emph{NuSTAR}'s FPMA and left to vary for FPMB and \emph{Swift}.
Spectra were rebinned to have at least 50 counts per bin. We started out with fitting an absorbed double blackbody ({\tt tbabs*(bbody+bbody)}) but the fit was poor, yielding a $\chi^{2}$ per degree of freedom (dof) of $1448/827$. The fit improved slightly by using an absorbed blackbody plus power-law ({\tt tbabs*(bbody+powerlaw)}), with $\chi^{2}$ over dof of 1152/827. Adding an additional component improved the fit significantly and therefore a more complex model (BB+double PL or BB+broken PL, $\chi^{2}$/dof=689/825 or 747/825, respectively) than suggested by previous data was favored. The F-test supports this conclusion, yielding a probability significantly smaller than $0.05$. All spectral fit results can be found in Table~\ref{tab:spectralresults}, while Figure~\ref{fig:bestfitspectra} shows our best-fit model for the spectra of \emph{NuSTAR} and \emph{Swift}.
%
\begin{table}
\footnotesize
\rotate
\begin{center}
\setlength{\tabcolsep}{1.5pt}
\caption{Best-fit parameters for 1E 2259+586 for phase-averaged and phase-resolved analysis.\label{tab:spectralresults}}
\vspace{2.0mm}
\begin{tabular}{lccccccccccc}
\tableline\tableline
Phase & Data\tablenotemark{a} & Energy & Fit Model\tablenotemark{b} & $N_{\rm H}\tablenotemark{c}$ & $kT$ & $\Gamma_{\rm s}/kT\tablenotemark{d}$ & $E_{\rm break}\tablenotemark{e}$ & $\Gamma_{\rm h}$\tablenotemark{f} & Flux\tablenotemark{g} & Ratio\tablenotemark{h} & $\chi^{2}/\rm dof$ \\ 
 & & [keV] & & $[\times 10^{22}$~cm$^{-2}]$ & [keV] & [-/keV] & [keV] & & & &  \tabularnewline
\tableline
$0$-$1$	& S     & $0.5$-$10.0$ & BB+BB   & $0.60(4)$ & $0.35(2)$   & $0.71(6)$ & $\cdots$      & $\cdots$     & $1.9(2)$            & $1.00(1)$   &   $331/537$  \\
$0$-$1$	& S     & $0.5$-$10.0$ & BB+PL   & $1.10(6)$ & $0.42(2)$   & $4.1(1)  $ & $\cdots$      & $\cdots$     & $2.1(1)$            & $1.71(9)$   &   $402/537$  \\
$0$-$1$	& S/N & $0.5$-$79.0$ & BB+BB   & $0.60(4)$ & $0.423(3)$ & $1.78(4)$ & $\cdots$      & $\cdots$     & $2.25(5)$          & $0.2(1)$     & $1448/827$  \\
$0$-$1$	& S/N & $0.5$-$79.0$ & BB+PL   & $1.10(6)$ & $0.38(1)$   & $3.91(3)$ & $\cdots$      & $\cdots$     & $2.2(1)$            & $2.5(3)$     & $1152/827$  \\
$0$-$1$	& S/N & $0.5$-$79.0$ & BB+2PL & $1.10(6)$ & $0.410(7)$ & $4.34(3)$ & $\cdots$      & 0.4(1)      & $2.36(7)$          & $0.31(2)$   &   $689/825$  \\
$0$-$1$	& S/N & $0.5$-$79.0$ & BB+bPL & $1.10(6)$ & $0.409(9)$ & $4.08(3)$ & $11.5(3)$ & $1.2(1)$  & $3.0(1)$            & $2.2(2)$     &   $747/825$  \\
\tableline
Peak A     & S/N & 0.5-79.0 & BB+2PL & $1.10(6)$ & 0.416(6) & 4.57(7)  & $\cdots$         & 1.0(3)  & 0.68(3) & 0.29(4) &    716/803  \\    
Peak B     & S/N & 0.5-79.0 & BB+2PL & $1.10(6)$ & 0.399(5) & 4.41(5)  & $\cdots$         & 0.3(1)  & 0.73(3) & 0.53(4) &    645/812   \\    
Off-pulse& S/N & 0.5-79.0 & BB+2PL & $1.10(6)$ & 0.404(6) & 4.87(9)  & $\cdots$         & 1.0(4)  & 0.48(3) & 0.37(6) &    879/1031  \\   
Peak A     & S/N & 0.5-79.0 & BB+bPL & $1.10(6)$ & 0.397(5) & 4.37(4) & 10.8(5)   & 1.38(1) & 1.51(4) & 0.78(2) & 776/803   \\    
Peak B     & S/N & 0.5-79.0 & BB+bPL & $1.10(6)$ & 0.386(5) & 4.26(5) & 10.9(4)  & 0.9(2)  & 1.46(2) & 0.85(1) & 696/812      \\    
Off-pulse& S/N & 0.5-79.0 & BB+bPL & $1.10(6)$ & 0.393(5) & 4.66(5) & 9.6(4)   & 1.4(4)  & 1.43(3) & 0.47(1) & 911/1031     \\    
\tableline
\end{tabular}
\tablenotetext{a}{\emph{NuSTAR} $5-79$~keV (N), \emph{Swift} $0.5-10$~keV (S)}
\tablenotetext{b}{Blackbody (BB), power-law (PL), broken power-law (bPL)}
\tablenotetext{c}{$N_{\rm H}$ was frozen to the value obtained from the phase-averaged spectral analysis for the phase-resolved analysis.}
\tablenotetext{d}{Soft photon index $\Gamma_{\rm s}$ if BB+PL model is used. Blackbody temperature of second (hot) BB if BB+BB is fit.}
\tablenotetext{e}{Break energy for the broken power-law (where applicable).}
\tablenotetext{f}{Photon index of the hard power-law component.}
\tablenotetext{g}{Unabsorbed flux in units of $10^{-11}$~erg cm$^{-2}$ s$^{-1}$ from $2-10$~keV for \emph{Swift} and $2-79$~keV if \emph{NuSTAR} data are included for all model components (BB+BB, BB+PL, BB+bPL) or two PL components (BB+2PL).}
\tablenotetext{h}{Ratio of flux in the $2-10$~keV range for \emph{Swift} and $2-79$~keV range if \emph{NuSTAR} is included. We show $F_{\rm BB(\rm hot)}/F_{\rm BB(\rm cold)}$ (BB+BB), $F_{\rm PL}/F_{\rm BB}$ (BB+PL), $F_{\rm bPL}/F_{\rm BB}$ (BB+bPL) and $F_{\rm PL(\rm hard)}/F_{\rm PL(\rm soft)} (BB+2PL)$.  }
\tablecomments{Numbers in parentheses indicate the $1 \sigma$ uncertainty in least significant digit. Cross-normalization factors were used when data from \emph{NuSTAR} and \emph{Swift} were combined. These factors were set to  $1.0$ for module A of \emph{NuSTAR} and to $1$ for \emph{Swift} if no \emph{NuSTAR} data were included. Cross-normalization factors were frozen to values obtained for phase-averaged spectral analysis, when performing phase-resolved spectral analysis simultaneously for \emph{NuSTAR} and \emph{Swift}. Fluxes are unabsorbed fluxes measured using XSPEC's {\tt cflux} model.}
\end{center}
\end{table}
%
%
%
%
%
\subsubsection{Phase-Resolved and Pulsed Spectral Analysis}\label{phaseresolved}
In this Section we present results from a phase-resolved spectral analysis based on the \emph{NuSTAR} and \emph{Swift} data to study the double-peak structure of the pulse profiles shown in Figure~\ref{fig:pulseprofile}. The phase intervals containing peak A and peak B were chosen to be $0.80-1.16$ and $1.16-1.52$, respectively. The off-peak spectrum covers the range of phase $0.52-0.80$. Before fitting the spectra we used {\tt grppha} to regroup the data such that there are at least 50 counts per spectral bin. We also use the same cross-normalization factors as obtained for the phase-averaged spectroscopy (\emph{NuSTAR}: 1.0 for FPMA, 1.06 for FPMB; \emph{Swift}: 0.94) and froze them for this fit along with the $N_{\rm H}$ value obtained above for the BB+PL model ($N_{\rm H}=1.10(6)\times 10^{22}$~cm$^{-2}$). To compare the spectra of the three phase intervals, we fit them with an absorbed blackbody plus a broken power-law model or a blackbody plus a double power-law. All fit results are included in Table~\ref{tab:spectralresults}. Some of the spectral parameters differ slightly for the three chosen phase ranges: the photon indices are slightly harder for peak B than they are for peak A and the softest power-law is obtained for the off-peak spectra. The blackbody temperatures for all three regions are consistent and the break point for the energy of the broken power-law fit is located around $10$~keV.

We also investigated the pulsed spectrum by subtracting the unpulsed emission from the total emission, but the statistics are too low to extract conclusive results. We cannot rule out a simple power-law or confirm a hardening at high energies (above $\sim15$~keV) as suggested by Kuiper et al. (2006). 
%
%
\subsubsection{Spectral Fits with the Coronal Outflow Model}\label{Beloborodov}
Similar to what was done by An et al. (2013) for \emph{NuSTAR} observations of 1E~1841$-$045, 
we apply the coronal outflow model of Beloborodov (2013a) to the observations of 1E 2259+586.
In this model the hard X-ray component is produced by a decelerating $e^{\pm}$ outflow in a twisted magnetic loop (the `j-bundle'). The relativistic $e^{\pm}$ flow is produced near the star with a Lorentz factor $\gamma \ga 10^3$ 
via continual discharge. As the outflow expands and fills the j-bundle, it experiences a radiative drag (deceleration) due to resonant scattering of thermal photons (with energies in the keV range) around the neutron star. Beloborodov (2013a,b) showed that the Lorentz factor of the flow decreases proportionally to the local magnetic field 
%
\begin{equation}
\label{eq_lflaw}
\gamma \sim 100 \frac{B}{B_Q},
\end{equation}
%
where $B_Q = m_e^2 c^3 / \hbar e \simeq 4.44 \times 10^{13}$ G.
Close to the star, where $B\ga 10^{13}$ G, the energy of scattered photons,
%
\begin{equation}
\label{eq_phnrj}
E_{\rm sc} \sim \gamma (B/B_Q) m_e c^2 \sim 50 (B/B_Q)^2 \ \mathrm{MeV} \sim 5 \gamma^2 \ \mathrm{keV} \, ,
\end{equation} 
%
is large enough to immediately convert to $e^{\pm}$ pairs. At larger radii, where $B \la 10^{13}$ G, the scattered photons escape, and the flow radiates almost all its kinetic energy before it reaches the magnetic equator, where the $e^{\pm}$ pairs annihilate. The predicted hard X-ray spectrum has an average photon index of $-1.5$ (this follows from Equations \ref{eq_lflaw} and \ref{eq_phnrj}) and cuts off in the MeV band. The hard X-rays are beamed along the magnetic field lines, and the observed spectrum varies greatly depending on the line of sight (see Figure 7 in Beloborodov 2013b).

To fit the \emph{NuSTAR} data, we make the simple assumption of an axisymetric j-bundle. The poloidal magnetic field is assumed to be close to the dipole configuration, and the magnetic dipole moment is fixed to the value inferred from spindown, $\mu_{\rm sd} \simeq 5.9 \times 10^{31}$ G. The model is described by 5 parameters:
(1) the angular position $\theta_j$ (magnetic colatitude) of the j-bundle footprint,
(2) the angular width $\Delta \theta_j$ of the  j-bundle footprint, 
(3) the power $L$ of the $e^{\pm}$ outflow along the j-bundle, 
(4) the angle $\alpha_{\rm mag}$ between the rotation axis and the magnetic axis, and 
(5) the angle $\beta_{\rm obs}$ between the rotation axis and the observer's line of sight. 
In addition, the reference point of the rotational phase, $\phi_0$, must be introduced as a free parameter when fitting the phase-resolved spectra. Note that $\Delta\theta_j=\theta_j$ would describe a polar cap. In this paper, we allow the j-bundle to have a ring-shaped footprint, which corresponds to $\Delta\theta_j<\theta_j$.

\begin{figure}[ht!]
\begin{centering}
\includegraphics[width=0.55\textwidth]{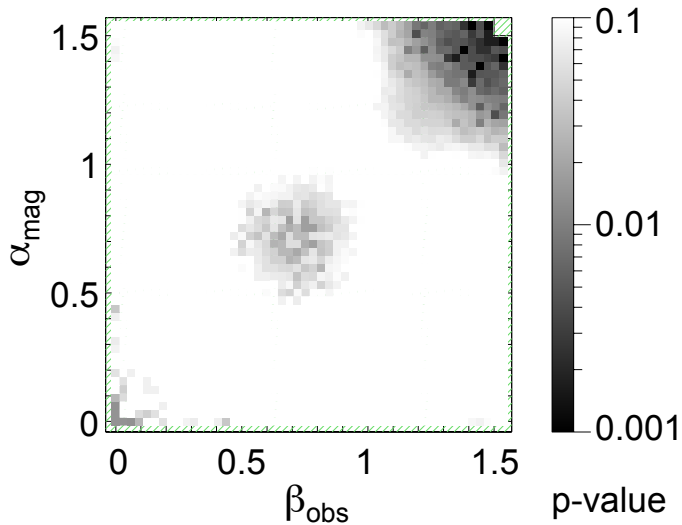}
\caption{Map of p-values for the fit of the hard X-ray component with the coronal outflow model. The p-values are shown in the plane of ($\alpha_{\rm mag}$, $\beta_{\rm obs}$) and maximized over the other parameters. The p-value scale is shown on the right. While the hatched green region has p-values smaller than $0.001$, the white region has p-values greater than $0.1$. Low p-values show parameters which are statistically disfavored (or excluded). Interchanging the values of $\alphamag$ and $\betaobs$ does not change the model spectrum, as long as the j-bundle is assumed to be axisymmetric. Therefore, the map of p-values is symmetric about the line of $\betaobs=\alphamag$.}
\label{fig:pvalue}
\end{centering}
\end{figure}

To test observations against the model, we follow the two-step method proposed by Hasco\"et et al. (2014).
We first explore the whole parameter space by fitting the phase-averaged spectrum of the total (pulsed+unpulsed) emission
together with three phase-resolved spectra of the pulsed emission. We only consider data above 16 keV where the observed hard component starts to dominate over the soft component as apparent from Figure~\ref{fig:bestfitspectra}. The results are shown in Figure~\ref{fig:pvalue}. We find that the model can fit the data. However, the fit does not give strong constraints on the parameters of the model; a large region in the $\alpha_{\rm mag}-\beta_{\rm obs}$ plane is allowed. The only significant constraint derived from this analysis is that the polar-cap shape of the j-bundle footprint does not
give a good fit; we find that a ring-shaped footprint is statistically preferred with $0.4 \la \theta_j \la 0.75$ and $\Delta \theta_j / \theta_j \la 0.2$ (at the $1\sigma$ level).

At the second step, we freeze the best-fit parameters for the outflow model
and fit the phase-averaged spectrum in the $0.5-79$ keV range using both \emph{NuSTAR} and \emph{Swift} data.
We test different models for the soft component,
taking into account the extension of the outflow contribution to low energies. Among these models BB$_{\rm tail}$ is a phenomenological modification of the blackbody (BB), 
where the Wien tail is replaced by a power-law smoothly connected to the Planck distribution (see Hasco\"et et al. 2014).
The energy $E_{\rm tail}$, at which the power-law starts, is a free parameter.
All results are summarized in Table \ref{tab_soft_comp}.
%
\newcommand{\markd}{\tablenotemark{a}}
\begin{table*}[t]
\begin{center}
\setlength{\tabcolsep}{1.5pt}
\caption{Parameters of the best-fit model for the soft X-ray component (\emph{NuSTAR} and \emph{Swift} phase-averaged spectra).}
\label{tab_soft_comp}
\vspace{4mm}
\footnotesize{
\begin{tabular}{lcccccccccc} \hline\hline
      Model     & $N_{\rm H}$      	          & $kT_{1}$ & $kT_{2}$	  &  $E_{\rm tail}$ & $L_1$\tablenotemark{a}  &  $L_2$\tablenotemark{a}  & $\Gamma$ &  $\nu L_{\nu}$\tablenotemark{a}  & $\chi^2$/dof & p-value\\ 
                     & ($10^{22}\ \rm cm^{-2}$) 	&  (keV)	   &  (keV)	           &	(keV)         &                                    &	      &        &  @ $1$ keV & & \\ \hline
 BB+BB               			&  0.44(2)  & 0.430(4)  & 1.22(2)     & $\cdots$ & 1.7(2)      & 0.073(3)  & $\cdots$ & $\cdots$ & 632/497 & $3.8\times 10^{-5}$ \\              
 BB+PL               			&  1.17(4)  & 0.43(1)    & $\cdots$  & $\cdots$ & 0.6(1)      & $\cdots$ & 4.28(5)    &  3.3(3)     & 526/497 & 0.18 \\              
 BB$_{\rm tail}$  			&  0.52(2)  & 0.401(6)  & $\cdots$  & $2.59(5)$& 1.13(9)    & $\cdots$ & $\cdots$ & $\cdots$ & 539/498  & 0.098 \\
 BB$_1$+ BB$_{\rm tail,2}$     &  0.62(4)  & 0.33(3)    & 0.65(9)     & $4.3(6)$  & 1.1(1)      & 0.3(2)      & $\cdots$ & $\cdots$ & 522/496 & 0.20 \\        
\\        
\hline
\end{tabular}}
\tablenotetext{a}{In units of $10^{35}(D/D_0)^2\ \rm erg\ s^{-1}$, where $D_0 = 4.0$~kpc is the distance inferred for 1E 2259+586~\cite{Tian2010}.}
\end{center}
\end{table*}
%
%
%
%
\section{Discussion}
\label{sec:Discussion}
In this paper we report on observations of 1E 2259+586 in X-rays obtained with the \emph{NuSTAR} and \emph{Swift} satellites. We observe a double-peak structure above $24$~keV for the first time at a statistical significance larger than $3\sigma$. Furthermore we find a hint of a phase shift towards lower phase values for one of the peaks in the pulse profile. We also find the pulsed fraction of 1E 2259+586 increases strongly with energy. To describe the phase-averaged total spectrum of the magnetar we use an absorbed blackbody plus a broken power-law. Furthermore, we test the coronal outflow model of Beloborodov (2013a) using the observed phase-averaged and phase-resolved spectra.
%
%
\subsection{Pulse Profile}
The \emph{NuSTAR} pulse profiles of 1E 2259+586 generally agree with those presented previously in Kuiper et al. (2006) based on \emph{RXTE} data. A change in peak predominance occurs with increasing energy at the boundary between the high-energy end of the soft component
(where thermal emission from the star is the dominant source) and the beginning of the hard component (where the $e^{\pm}$ outflow becomes the dominant source).

The two pulses of the double-peak profile seen in 1E 2259+586 are separated by half a period and have similar peak fluxes. This might be an indication that the neutron star is seen from the same magnetic latitude\footnote{While the rotation axis of the neutron star is fixed, the orientation of the magnetic axis relative to the observer changes with phase if $\alpha_{\rm mag}\ne 0$. As a result the magnetic latitude of the observer line of sight also changes with phase.} (in absolute value) every half period (assuming that the emission pattern around the neutron star is relatively axisymmetric). This is possible if (and only if) $\alpha_{\rm mag} \approx \pi /2$ (corresponding to a nearly orthogonal rotator) or $\beta_{\rm obs} \approx \pi /2$.  Note that no specific line of sight is \emph{a priori} favored and $\beta_{\rm obs} \approx \pi/2$  is the most probable observer angle. We also note that the center of pulse B shifts from $1.38$ to $1.35(2)$ for energies above $8$~keV, while the initially dominant peak A remains centered at the same phase independent of energy. The phase shift of peak B is, however, not significant with current statistics, and more data are needed to confirm or rule out the observed trend. The $e^{\pm}$ outflow and its emission are probably not perfectly axisymetric, which can easily result in an energy-dependent phase shift of the peak maxima. Modeling such a minor shift with a small number of parameters is difficult and would require a different, more complicated fitting procedure.

Both the area and the peak-to-peak pulsed fraction as well as $PF_{\rm RMS}$ show an increase with energy. The area pulsed fraction is about $60\%$ for energies around $10$~keV and rises to close to $100\%$ at $20$~keV, $PF_{\rm RMS}$ is around $30\%$ and $70\%$ at these energies, respectively. The \emph{Swift} and \emph{NuSTAR} values for area and peak-to-peak pulsed fraction are consistent with those reported by Patel et al. (2001) as peak-to-peak pulsed fraction for $E=0.5-7.0$~keV of $35.8\%\pm 1.4\%$. The use of different analysis techniques precludes a direct comparison of the pulsed fractions we determined and those established in some previous publication (spectral pulsed fraction of $\gtrsim 43\%$ for energies below $10$~keV as reported by Kuiper et al. 2006). We can compare, however, the general behavior for the pulsed fraction versus energy observed for 1E 2259+586 with that for other magnetars. Kuiper et al. (2006) reported an increase of pulsed fraction with energy for 1E 1841$-$045, 4U 0142+61 and 1RXS J1708$-$4009. An et al. (2013) observed the same tendency for 1E 1841$-$045, although the trend was not quite as pronounced ($24\%\pm4\%$ RMS pulsed fraction at $20$~keV and $41\%\pm18\%$ at $80$~keV). For 1E 2259+586 we clearly see a dramatic increase in pulsed fraction following the general trend observed for other magnetars. Such an increase in the pulsed fraction is naturally expected in the coronal outflow model: photons of higher energy are more strongly beamed along 
the magnetic field lines as they are produced where the flow moves with a higher Lorentz factor (see Equation 9).

%
%
\subsection{Spectrum}
We found that the spectral parameters of \emph{Swift} and \emph{NuSTAR} for the phase-averaged spectrum (see Section~\ref{sec:phaseaveragedspectralana}) agree well with those reported by Patel et al. (2001) and Zhu et al. (2008). These results support the fact that especially the soft-band spectrum (below $\sim10$~keV) has been stable over a period of $\sim13$~yrs from \emph{Chandra} observations in 2000~\cite{Patel2001} to 2013 despite an anti-glitch~\cite{Archibald2013} and glitches~\cite{Kaspi2003,Icdem2012}, that temporarily altered the spectral parameters.
%
\begin{figure}[t!]
\begin{center}
\begin{tabular}{cc}
\includegraphics[angle=90, width=0.49\textwidth]{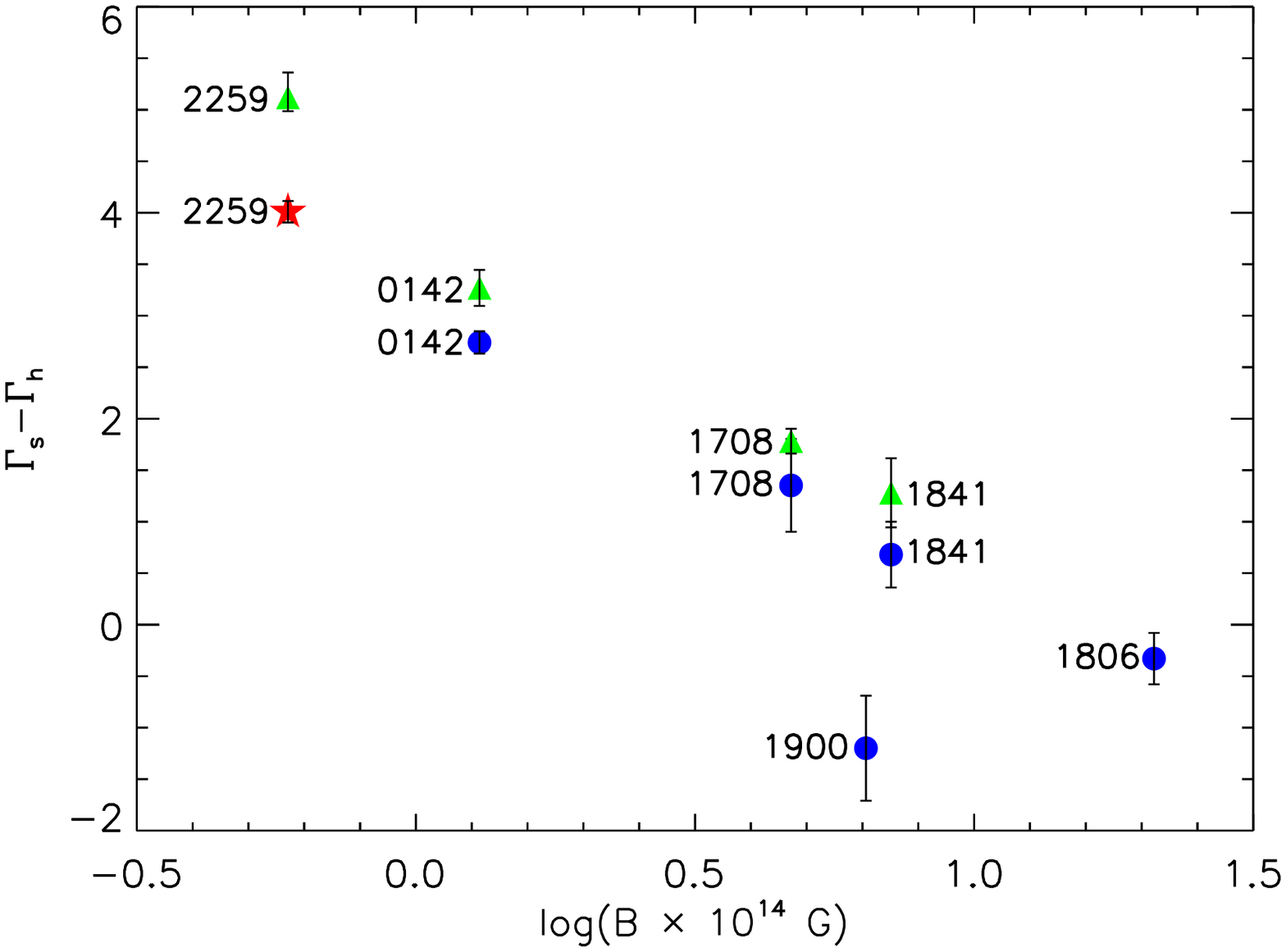}  & \includegraphics[angle=90,width=0.49\textwidth]{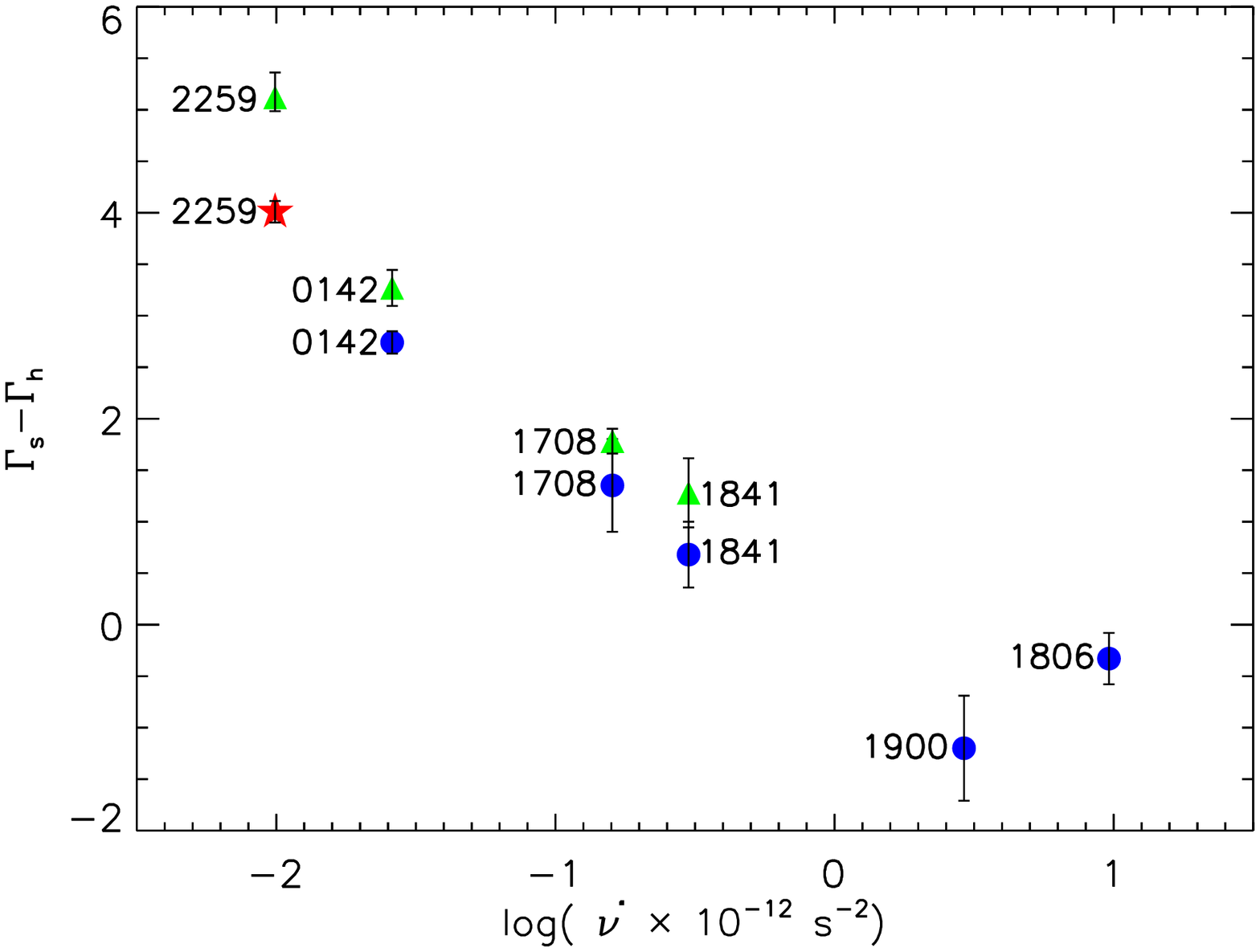} 
\end{tabular}
\end{center}
\caption{Spectral turnover ($\Gamma_{\rm s}-\Gamma_{\rm h}$) versus magnetic field $B$ (left) and spin-down $\dot{\nu}$ (right) from Kaspi \& Boydstun (2010) for all magnetars presented there as well as our new result. Blue circles represent total flux, green triangles represent pulsed flux. Our total flux result is indicated by the red star. $1\sigma$ uncertainties are shown. For details on full list of magnetar names, included data and references, see Kaspi \& Boydstun (2010).}
\label{fig:Boydstun}
\end{figure}
%

With the new data we were able to test if an additional power-law component is required and find that it is. Therefore 1E 2259+586 exhibits the general tendency for magnetars to get harder at X-ray energies above $10$~keV. We parametrized the \emph{NuSTAR} plus \emph{Swift} spectrum with a blackbody plus double power-law model as well as a blackbody plus a broken power-law. This phenomenological parameterization is a convenient way to characterize the presence of a hard X-ray component; the physical model of a coronal outflow 
is discussed in Section~\ref{sec:DiscussionBelo} below.

We note that the results of our analysis support the anticorrelation of spectral turnover ($\Gamma_{\rm s}-\Gamma_{\rm h}$) with magnetic field $B$ and spin-down $\dot\nu$ suggested by Kaspi \& Boydstun (2010). The authors demonstrated that there is a trend suggesting that those magnetars with the highest (lowest) $B$ or $\dot{\nu}$ have the smallest (largest) spectral turnover. In Figure~\ref{fig:Boydstun}, we summarize the results of Kaspi \& Boydstun (2010) for pulsed (green triangles) and total fluxes (blue circles) and add our new findings for 1E 2259+586 (red star) which further strengthens their case. Errors shown are $1\sigma$ uncertainties. It is interesting to point out that the observed spectral turnover for energetic rotation-powered pulsars (RPPs) is practically zero despite their much lower magnetic fields. Therefore Kaspi \& Boydstun (2010) concluded that the production mechanism for X-rays above $10$~keV in RPPs differs significantly from the one in magnetars. We also find a value for the hardness ratio, i.e. $F_{\rm h}/F_{\rm s}$ (flux ratio of hard to soft spectral component in the $2-79$~keV band), that agrees well with the observed correlation of hardness ratio and characteristic age that has been inferred from the spin-down rate, as reported by Enoto et al. (2010). They found a decrease in hardness ratio with increasing characteristic age or, likewise, an increase of $F_{\rm h}/F_{\rm s}$ with magnetic field $B$.

Furthermore, we investigate the phase-resolved spectra for \emph{NuSTAR} and \emph{Swift} and find that the PL indices were slightly harder for peak B than for peak A, while the breaking point for the energy is located around $10$~keV for both pulses. We also examine pulsed spectra extracted from the \emph{NuSTAR} observation, but due to low statistics we are not able to rule out a single power-law. More high-energy data are needed in order to compare the results to those of Kuiper et al. (2006).

We comment further that our analysis did not find any spectral lines or absorption features in the phase-averaged and phase-resolved spectra that could be interpreted as cyclotron lines~\cite{Tiengo2013}. We also searched for short bursts in the data, but in contrast to 1E 1048.1$-$5937, for which recently six bright X-ray bursts have been reported in \emph{NuSTAR} data~\cite{An2014}, we did not observe a similar behavior for 1E 2259+586.
%
%
\subsection{Spectral Modeling with the Coronal Outflow Model}
\label{sec:DiscussionBelo}
We find that the phase-resolved spectra of 1E 2259+586 are consistent with the model of Beloborodov (2013a), although the available data do not allow us to derive strong constraints on the parameters of the model. In contrast to 1E 1841$-$045~\cite{An2013}, 4U  0142+61, and 1RXS J1708$-$4009~\cite{Hascoet2014}, a broad range of $\alpha_{\rm mag}$ and $\beta_{\rm obs}$ is so far allowed for 1E 2259+586. The reason for this degeneracy is twofold: 
(i) the statistics of \emph{NuSTAR} data of 1E 2259+586 are not as good as for the other objects; 
(ii) the magnetic dipole moment of 1E 2259+586 is rather low, one order of magnitude smaller than for 1E 1841$-$045.
As a result, the boundary of the radiative zone (where $B\sim 10^{13}$ G) is closer to the star. This leads to a larger allowed range of the j-bundle latitudes and to more flexible predictions of the model. Better statistics (with longer exposure times), 
extension of observations to the MeV range (where the emission of the $e^{\pm}$ outflow peaks),
or polarization measurements would help to break the degeneracy of $\alpha_{\rm mag}$ and $\beta_{\rm obs}$.

The footprint location of the j-bundle is better constrained, with $0.4 \la \theta_j \la 0.75$ and $\Delta \theta_j / \theta_j \la 0.2$ (at the $1\sigma$ level).
This means that a ring-like footprint located at rather high colatitudes\footnote{Complementary angle of the latitude, i.e. difference between $90\arcdeg$ and latitude.} is statistically preferred to a polar cap footprint covering low colatitudes. This contrasts with the results obtained for 1E 1841$-$045 \citep{An2013}, 4U 0142+61, and 1RXS J1708$-$4009 \cite{Hascoet2014}, where a polar cap footprint provided a good fit.
If confirmed with future, higher-statistics data, it would point to diversity in the crustal motions responsible for the twist of the magnetosphere.

We also analyzed the soft component, taking into account the extension of the outflow emission to low energies.
The results are not very sensitive to the choice of parameters for the coronal outflow, 
because all good fits have similar extensions to low energies, which cut off below $\sim 5$~keV. We find that the models BB+PL and BB+BB$_{\rm tail}$ provide equally good fits to the soft component; the fit with a modified blackbody (BB$_{\rm tail}$) is only slightly worse but still acceptable, while the two-blackbody model (BB+BB) is statistically unacceptable.
Even though BB+PL provides a good fit, it is physically problematic because the power-law 
component is brighter than the blackbody (at all frequencies) and its energy content diverges at the low energy end of the spectrum (in case of no cut off). It likely overpredicts the (sub) keV flux and the hydrogen column density $N_{\rm H}$. Our physically preferred model is BB+BB$_{\rm tail}$, where the cold blackbody corresponds to emission from most of the neutron star surface and the hot modified blackbody to a hot spot on the star (Hasco\"et et al. 2014). Such a spot may be expected at the footprint of the j-bundle, as some particles produced in the $e^\pm$ discharge bombard the footprint and heat it (Beloborodov 2009). The area of the cold and hot thermal components are $\mathcal{A}_1 \approx 0.7$~$ \mathcal{A}_{\rm NS}$ and $\mathcal{A}_2 \approx 0.02$~$\mathcal{A}_{\rm NS}$,  respectively, where $\mathcal{A}_{\rm NS}$ represents the surface area of a neutron star with a typical radius of $R_{\rm NS} = 10$ km. These results are similar to those obtained for 1E 1841$-$045, 4U  0142+61, and 1RXS J1708$-$4009 (Hasco\"et et al. 2014). Note however that `BB$_{\rm tail}$ only' also provides an acceptable fit for the soft component in 1E 2259+586, and the parameters of the hot spot in this 
source are not strongly constrained from these data.
%
\section{Conclusion}
We present an analysis of simultaneous \emph{NuSTAR} and \emph{Swift} observations of the AXP 1E 2259+586, and report on our spectral and temporal results.
We find that the double-peak pulse profile in different energy bands generally agrees with previous \emph{RXTE} results, indicating a gradual change in peak dominance with energy. We used Fourier analysis techniques to quantify the similarities and differences and also show for the first time pulsations in the energy band above $24$~keV for non-background-subtracted \emph{NuSTAR} data. There is a hint at a change of position in phase for one of the two peaks in the pulse profile with increasing energy, but more data are needed to confirm this observation. This might suggest the pulsar is not a nearly orthogonal rotator. We also present the area and peak-to-peak pulsed fractions along with the RMS variation, showing that they all increase with energy and reach $96\%\pm14\% $ ($PF_{\rm area}$) and $71\%\pm15\%$ ($PF_{\rm RMS}$) at around $20$~keV, similar to what is seen in other magnetars~\cite{Kuiper2006}. The phase-averaged spectral analysis for the magnetar finds that there is good agreement between previous \emph{Chandra} and \emph{RXTE} data and the new observations. We also report that a model with an addition PL component (BB+2PL or BB+broken PL) is statistically preferred for the pulsar spectrum including the hard X-ray \emph{NuSTAR} data. Our result for parametrization of the spectral turnover supports the anticorrelation hypothesis of Kaspi \& Boydstun (2010) and we obtain a value of  $\Gamma_{\rm s}-\Gamma_{\rm h}=4.0(1)$ for $B=0.59\times 10^{14}$~G. Our findings also agree with the observed correlation of hardness ratio and characteristic age as reported by Enoto et al. (2010). The observed phase-resolved spectra are consistent with the coronal outflow model of Beloborodov (2013a). The fits of the spectra by the model prefer a ring-like j-bundle, 
in contrast to 1E 1841$-$045, 4U  0142+61, and 1RXS J1708$-$4009 (Hasco\"et et al. 2014). Unfortunately the available
data do not provide significant constraints on the magnetic inclination ($\alpha_{\rm mag}$) or the observer line of sight ($\beta_{\rm obs}$). 

\acknowledgments

Part of this work was performed under the auspices of the U.S. Department of Energy by Lawrence Livermore National Laboratory under Contract DE-AC52-07NA27344 with support from the LDRD program through grant 13-ERD-033. This work was supported under NASA Contract No. NN\-G\-0\-8\-F\-D60C, and made use of data from the \emph{NuSTAR} mission, a project led by the California Institute of Technology, managed by the Jet Propulsion Laboratory, and funded by the National Aeronautics and Space Administration. We thank the \emph{NuSTAR} Operations, Software and Calibration teams for support with the execution and analysis of these observations. This research has made use of the \emph{NuSTAR} Data Analysis Software (NuSTARDAS) jointly developed by the ASI Science Data Center (ASDC, Italy) and the California Institute of Technology (USA). VMK receives support from an NSERC Discovery Grant and Accelerator Supplement, from the Centre de Recherche en Astrophysique du Qu\'ebec, an R. Howard Webster Foundation Fellowship from the Canadian Institute for Advanced Study, the Canada Research Chairs Program and the Lorne Trottier Chair in Astrophysics and Cosmology. AMB is supported by the NASA ATP grant NNX 13AI34G. This work made use of data supplied by the UK Swift Science Data Centre at the University of Leicester. We also thank Dr. A. M. Archibald for helpful discussions.\\

\bibliographystyle{apj}

\clearpage

\end{document}